\documentclass[twoside]{article} 
\usepackage[accepted]{aistats2014}
\usepackage{hyperref}
\usepackage{url}
\usepackage{amsmath}
\usepackage{enumerate}
\usepackage{graphicx} 
\usepackage{amssymb}
\usepackage{natbib}

\newcommand{\RR}{I\!\!R}
\newcommand{\rd}{{\rm d}}
\newcommand{\thetahat}{\hat{\theta}}
\newcommand{\piG}{\pi_{\mbox{{\tiny GABC}}}}
\newcommand{\piGhat}{{\hat\pi}_{\mbox{{\tiny GABC}}}}

\newcommand{\BE}{{\mathbb{E}}}
\newcommand{\CD}{{\mathcal{D}}}
\newcommand{\CN}{{\mathcal{N}}}

\begin{document}
\bibpunct{(}{)}{,}{}{}{}

\twocolumn[

\aistatstitle{Accelerating ABC methods using Gaussian processes}

\aistatsauthor{ Richard D. Wilkinson}

\aistatsaddress{ School of Mathematical Sciences, University of Nottingham,
Nottingham, NG7 2RD, United Kingdom } 
]



\begin{abstract}

Approximate Bayesian computation (ABC)  methods are used to approximate  posterior distributions  using simulation rather than likelihood calculations. We introduce Gaussian process (GP) accelerated ABC, which we show can significantly reduce the number of simulations required. As  computational resource is usually the main determinant of accuracy in ABC, GP-accelerated methods can thus enable  more accurate inference in some models.  GP models of the unknown log-likelihood function are  used to exploit continuity and smoothness,  reducing the required computation. We use a sequence of models  that increase in accuracy, using intermediate models to rule out regions of the parameter space as implausible. The methods will not be suitable for all problems, but when they can be used, can result in significant computational savings. For the Ricker model, we are able to achieve accurate approximations to the posterior distribution using a factor of 100 fewer simulator evaluations than comparable Monte Carlo approaches, and for a population genetics model we are able to approximate the exact posterior for the first time.

\end{abstract}

\section{Introduction}

Approximate Bayesian computation (ABC) is the term given to a collection of algorithms used for calibrating complex simulators \citep{Csillery_etal2010, Marin_etal2012}. Suppose $f(\theta)$ is a simulator that models some physical phenomena for which we have observations $D\in\RR^d$, and that it takes unknown parameter value $\theta \in \RR^p$ as input and returns output $X\in\RR^d$. 
The Bayesian approach to calibration is to find the posterior distribution 
$ \pi(\theta | D) \propto \pi(D|\theta) \pi(\theta)$,
where $\pi(\theta)$ is the prior distribution  and $\pi(D|\theta)$ is the likelihood function defined by the simulator. 

ABC algorithms enable the posterior to be approximated using realizations from the simulator, i.e., they do not require knowledge of $\pi(D| \theta)$. They have become popular in a range of application areas, primarily in the biological sciences \citep{Beaumont_etal02, Toni_etal10, Beaumont_2010}. This popularity stems from their universality (it is nearly always possibly to use some form of ABC algorithm) and their simplicity (complex likelihood calculations are not required). The simplest ABC algorithm is based on the rejection algorithm:
\begin{enumerate}
 \item Draw $\theta$ from the prior: $\theta \sim \pi(\theta)$
\item Simulate a realization from the simulator: $X \sim \pi(X|\theta)$
\item Accept $\theta$ if and only if $\rho(D,X)\leq \epsilon$ 
\end{enumerate}
where $\rho(\cdot, \cdot)$ is a distance measure on $\RR^d$. The tolerance, $\epsilon$, controls the trade-off between computability and accuracy. When $\epsilon=\infty$ the algorithm returns the  prior distribution. Conversely, when $\epsilon=0$ the algorithm is exact and gives draws from $\pi(\theta|D)$, but acceptances  will be rare.

Accuracy considerations dictate that we want to use a tolerance value as small as possible, but computational constraints limit what is feasible, and it is dealing with this limited computational resource that is the key challenge for ABC methods.  If the simulator output is complex, then for small tolerance values (and thus high accuracy) the simulator output will rarely be close enough to the observations, and we will thus require a large number of simulator runs to generate sufficient accepted parameter values to approximate  the posterior. Even if the simulator is computationally cheap, an ABC algorithm may still require many hours of computation to approximate a posterior distribution, even to moderate accuracy. 
Extensive work has been done on developing  algorithms that more efficiently explore the parameter space than rejection-ABC. There are ABC versions of MCMC \citep{Marjoram_etal03}, sequential Monte Carlo (SMC) \citep{Sisson_etal07, Toni_etal08}, and many other Monte Carlo algorithms \citep{Marin_etal2012}.
These algorithms all share the following properties:
 (i) They sample space randomly and only learn from previous simulations in the  limited sense of using the current parameter value to determine which move to make next;
(ii) They do not exploit known properties of the likelihood function, such as continuity or smoothness.
These properties  guarantee the asymptotic success of the algorithms. However, they also make them computationally expensive as the algorithm has to learn details that were known a priori, for example, that the posterior density is a smooth continuous function.

In this paper we use Gaussian process (GP) models of the likelihood function to accelerate ABC methods and thus enable more accurate inference given limited computational resource. The approach can be seen as a natural extension of the synthetic likelihood approach proposed in  \citet{Wood_2010} and the implicit  inference approach of \citet{Diggle_etal84}, and follows the example of  \citet{Rasmussen2003} who used GPs to accelerate hybrid Monte Carlo methods.
We use space filling designs rather than random sampling,  and use the idea of sequential history matching \citep{Craig_etal1997, Vernon_etal2010} to successively rule out swathes of the parameter space as implausible. We are thus able  to build  accurate  models of the log-likelihood function that can be used to find the posterior distribution using far fewer simulator evaluations than is necessary with other ABC approaches.

\section{GP models of the ABC likelihood}


\citet{Wilkinson_2013} showed that any ABC algorithm gives Monte Carlo exact inference, but for a different model to  the one intended. If we replace step 3 in the rejection algorithm  by {\it `Accept $\theta$ with probability proportional to $\pi(D|X)$'}, 
where $\pi(D|X)$ is an acceptance kernel, we get a generalized ABC (GABC) algorithm. If we make the choice $\pi(D|X) \propto I\!\!I_{\rho(D,X) \leq \epsilon}$, then we are returned to the uniform rejection-ABC algorithm above. The GABC algorithm gives a Monte Carlo exact approximation to 
$$\piG(\theta|D) = \frac{\int \pi(D|X) \pi(X|\theta) \rd X \pi(\theta)}{\pi(D)}$$
where we can interpret this as the posterior distribution for the parameters when we believe $\pi(D|X)$ represents a statistical model relating the simulator output to the observations. For example, $\pi(D|X)$ might model a combination of measurement error on the observations and the  simulator discrepancy. The special case  $\pi(\rd D|X)=\delta_X(\rd D)$, i.e., a  point mass at $X$, represents the situation where we believe the simulator is a perfect model of the data, and gives the posterior distribution  $\pi(\theta|D) \propto \pi(D|\theta) \pi(\theta)$. 

We can approximate the GABC likelihood  function $\piG(D|\theta) = \int \pi(D|X) \pi(X|\theta) \rd X$ by the unbiased Monte Carlo sum
\begin{equation}\label{eqn:piGhat}
\piGhat (D|\theta) = \frac{1}{M} \sum_{I=1}^M \pi(D|X_i)
\end{equation}
where $X_1, \ldots, X_M \sim \pi(X | \theta)$, and by repeating for different $\theta$ we can begin to build a model of the likelihood surface as a function of $\theta$.
The idea is  related to the concept of emulation  \citep{Kennedy_etal01}, but whereas they emulate the simulator output (possibly a high dimensional complex function), we  instead emulate the GABC likelihood function (a one dimensional function of $\theta$). 

Often in ABC algorithms a summary function $S(\cdot)$ is used to project the data and simulations into a lower dimensional space, and then 
instead of finding $\piG(\theta |D)$ we instead find $\piG(\theta | S(D))$, i.e., the posterior for $\theta$ based on the summary statistics, rather than the full data (Equation (\ref{eqn:piGhat}) becomes the sum of $\pi(S(D) | S(X))$ terms). \citet{Wood_2010} proposed a related approach, and  assumed  a Gaussian synthetic likelihood function 
\begin{equation}\label{eqn:WoodLikelihood}
\piGhat(S(D)|\theta)=\phi(S(D); {\hat \mu}_\theta, {\hat \Sigma}_\theta) 
\end{equation}
where $\phi$ is the multivariate Gaussian density function and ${\hat \mu}_\theta$ and ${\hat \Sigma}_\theta$ are the mean and covariance of $S(X)$ estimated from the $M$ simulator evaluations at $\theta$. \citet{Drovandi_etal2014} have recently reinterpreted this  as a Bayesian indirect likelihood (BIL) algorithm, and drawn links with indirect inference (ABC-II). In BIL and ABC-II a tractable auxiliary model for the data is proposed, $p(D|\psi)$ say, and simulations from the model run at $\theta$ are used to estimate $\psi$ for the auxiliary model. The approach demonstrated here is to learn the mapping from $\theta$ to $\psi$ in the specific case of \citet{Wood_2010}, but can  be used to  accelerate ABC-II and BIL approaches more generally.

\subsection{Gaussian process models of the likelihood}

For many simulators and choice of acceptance kernel, if not the majority, the GABC likelihood function will be a smooth continuous function of $\theta$. 
If so, then the value of $\piG(D|\theta)$ is informative about $\piG(D| \theta+h)$ for small $h$. This allows us to model  $\piG(D|\theta)$ as a function of $\theta$. Other ABC methods and the approach of \citet{Wood_2010}, do not assume continuity of the likelihood function, and the algorithms  estimate $\piG(D|\theta)$ and $\piG(D|\theta+h)$ independently. Moreover, if the algorithm returns to $\theta$ (for example in an MCMC chain), they generally re-estimate $\piG(D|\theta)$ despite having a previous estimate available. 

We aim to reduce the number of simulator evaluations required in the inference by using a   model of the unknown likelihood function. Once the model has been trained and tested, we can then use it to calculate the posterior.
The likelihood function is difficult to work with as it varies from 0 to very small values, and is required to be positive. We instead model the log-likelihood $l(\theta) = \log \piG(D|\theta)$ which we estimate\footnote{To avoid numerical underflow, we use the log-sum-exp trick $\log\sum e^{a_i} = \log \sum e^{a_i-A} + A$, where $e^{a_i}=\pi(D|X_i)$ and $A=\max a_i$.} by $\hat{l}_M(\theta)=\log \piGhat (D|\theta)$. We model $\l(\cdot)$ as a Gaussian process and
assume  a priori that $l(\cdot) \sim GP(m_\beta(\cdot), c_\psi(\cdot, \cdot))$  where $m_\beta(\cdot)$ and $c_\psi(\cdot, \cdot)$ are the prior mean and covariance functions respectively \citep[see][for an introduction to GPs]{Rasmussen_etal2006}. 
For some models, using  a linear model for the mean function of the form
$$m_\beta(\theta)=\beta_0 + \theta^T \beta_1 + \operatorname{diag}(\theta \theta^T) \beta_2,$$
provides more accurate results with fewer situations. The  quadratic term  is included as we expect $l(\theta) \rightarrow -\infty$ as $\theta \rightarrow \pm \infty$, and so inclusion of $\theta^2$  improves the prediction of the GP when extrapolating outside of the design region. More complex mean functions are used on a problem specific basis, with the choice guided by diagnostics plots.

We use a covariance function of the form
$$c_\psi(\theta_i, \theta_j) = \tau^2 c_\lambda(\theta_i, \theta_j) + v^2I\!\!I_{i=j}$$
where $c_\lambda$ is usually taken to be of a standard form such as a squared exponential or Mat\'ern covariance function, with a vector of length scales, $\lambda$, that needs to be estimated. The nugget term is included because  $\hat{l}_M(\theta_i)$ are noisy observations of the likelihood $l(\theta_i)$, with the nugget variance, $v^2$, taken to be the sampling variance of $\hat{l}_M(\theta)$. We estimate $v^2$ by using the bootstrapped variance of the terms in the log-likelihood estimate, which helps  avoid non-identifiability in the estimation of the other GP parameters. 
We use a conjugate improper normal-inverse-gamma prior $\pi(\beta, \tau^2) \propto 1/\tau^2$ for these  parameters, which allows them to be  integrated out analytically and  use a plug-in approach  for the length-scale parameters, $\lambda$, estimating them using maximum likelihood.
The posterior distribution of the GP given the training ensemble $\mathcal{E}$ (see below), then has a multivariate t-distribution with updated mean and covariance functions $m^*(\theta)$ and $c^*(\theta, \theta')$. Details can be found in \citet{Rasmussen_etal2006}.

\subsection{Design}\label{sect:design}
To train the GP model, we use an ensemble $\mathcal{E} = \{(\theta_i, \hat{l}_N(\theta_i))_{i=1}^N\}$ of parameter values and estimated GABC log-likelihood values. The experimental design $\{\theta_i\}$ at which we evaluate the simulator is carefully chosen in order to minimize the number of design points (and thus the number of simulator evaluations) needed to achieve sufficient accuracy \citep{Santner_etal2003}. We use a $p$-dimensional Sobol sequence to generate an initial space filling design on $[0,1]^p$. This is a quasi-random low discrepancy sequence which uniformly fills space \citep{Morokoff_etal1994}. The advantage of Sobol sequences over other space filling designs, such as maxi-min Latin hypercubes, is that they can be extended when required. The use of quasi-random numbers in Monte Carlo sampling has been recently explored by \citet{Barthelme_etal2013} and \citet{gerber2014sequential}, who used low-discrepancy sequences to reduce  the Monte Carlo error  in (expectation-propagation) ABC and SMC.

To generate a design that fills the space defined by the prior support $\Theta_0 = \operatorname{supp}(\pi(\cdot))$, in a manner that places more points in the more (a priori) likely regions of space, we translate the Sobol design on $[0,1]^p$ into $\Theta_0$. If $\pi(\theta)$ is a product of uniform distributions, this can be done with a simple linear transformation. If the prior is non-uniform, but each parameter is a priori independent, then we apply the inverse cumulative density function (CDF) to each parameter. Depending on whether the prior is misspecified or not, it may be necessary to expand the design outwards, by inflating the variance used in the inverse CDFs. 

\section{Sequential history matching}

For most complex inference problems, this approach alone will not be sufficient as the log-likelihood often ranges over many orders of magnitude. For the Ricker model described in Section \ref{sect:Ricker}, the estimated log-likelihood varies from approximately $-5$  to $-10^3$ and most models will struggle to accurately model $l(\theta)$ over the entire input domain $\Theta_0$. However, only values of $l(\theta)$ within a certain distance of $l(\thetahat)$, where $\thetahat$ is the maximum likelihood estimator, are important for estimating the posterior distribution. If $\exp(l(\theta)+\log \pi(\theta))$ is orders of magnitude smaller than $\exp(l(\thetahat)+\log \pi(\thetahat))$, then the posterior density $\pi(\theta|D)$ will be approximately $0$. Thus, we do not need a model capable of accurately predicting $l(\theta)$, only one capable of predicting that $l(\theta)$ is small compared to $l(\thetahat)$.

We use the idea of sequential history matching \citep{Craig_etal1997} to  iteratively rule out regions of the parameter space as implausible (in the sense that the parameter could not have generated the observed data). We build a sequence of GP models, each of which is used to define regions of space that are implausible according to the criterion below. Models are then defined only on regions of space not already ruled implausible by the previous model in the sequence.

\subsection{Implausibility}

Suppose that  we have  a Gaussian process model $\eta(\cdot)$ of $l(\cdot)$ built using training ensemble $\mathcal{E}$, and that the prediction of $l(\theta)$ has mean $m$ and variance $\sigma^2$. We define $\theta$ to be implausible (according to $\eta$) if 
\begin{equation}\label{eqn:implausibility} 
m + 3\sigma < \max_{\theta \in \mathcal{E}} \hat{l}_M(\theta) - T,
\end{equation}
where $T>0$ is a threshold value, chosen so that if $l(\thetahat)-l(\theta) > T$, then $\pi(\theta | D)/\pi(\thetahat|D) \approx 0$ and $\theta$ can be discounted\footnote{Implausibility  is defined only for uniform prior    here, but can easily be extended to non-uniform distributions.}. 
The right-hand side of Equation (\ref{eqn:implausibility})  describes a log-likelihood value below which we believe the posterior will be approximately zero. The left-hand side is the GP prediction of $l(\theta)$ plus three standard deviations (so that the estimated probability of $l(\theta)$ exceeding $m+3\sigma$ is less than $ 0.003$). Thus,  the implausibility criterion rules out points for which the GP model gives only a small probability of the log-likelihood exceeding the threshold at which $\theta$ is important in the posterior. A point which is not ruled implausible by Equation (\ref{eqn:implausibility}), may still have a posterior density close to zero (i.e., it may not be plausible), but the GP model currently in use is not  able to rule it out.

\begin{figure*}
\begin{center}
\includegraphics[width=0.75\linewidth]{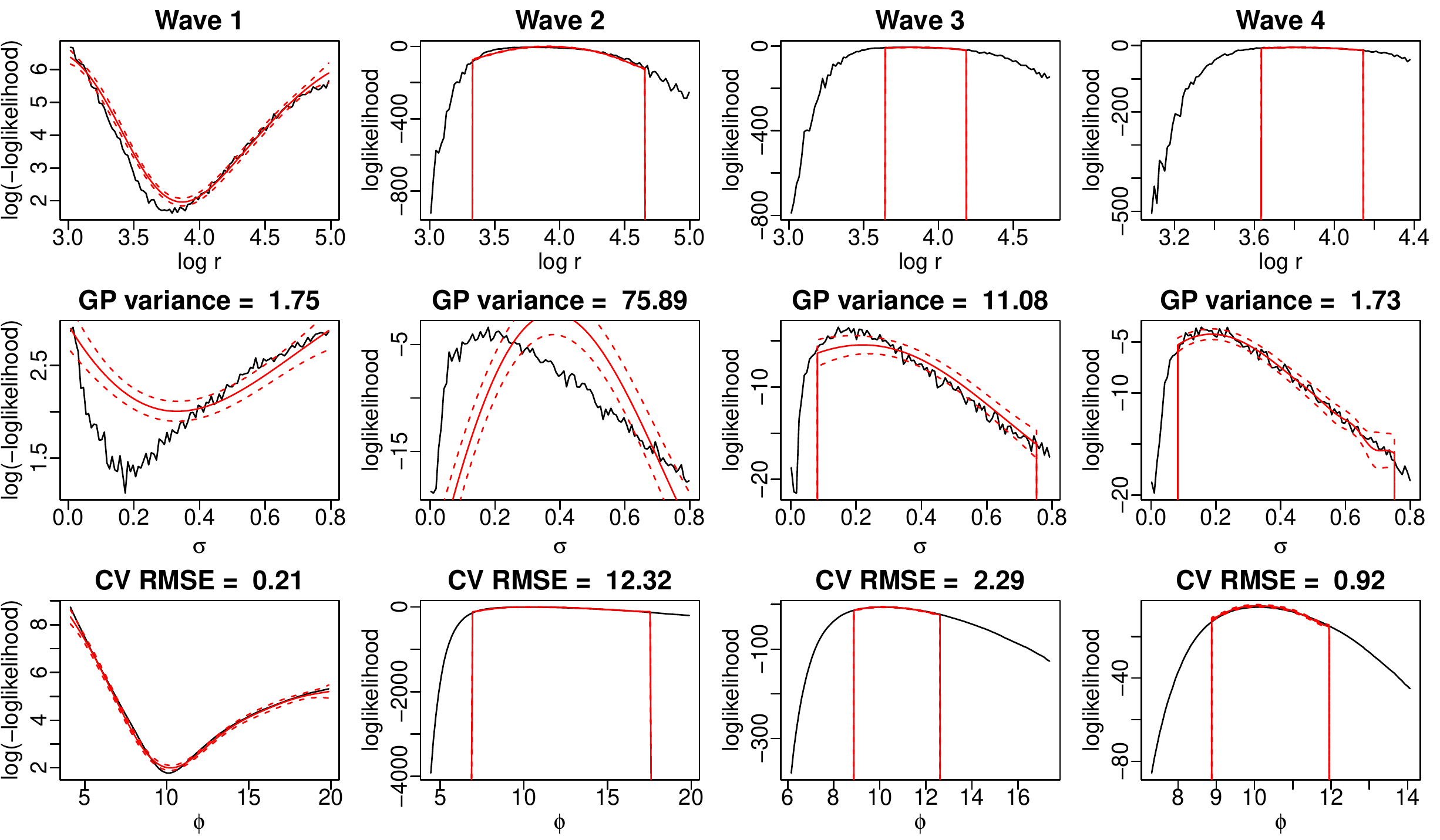} 
\end{center}
\caption{Diagnostics plots for the fitted GPs with one column per wave. A slice through the likelihood surface at  $\theta=(3.8,   0.3,  10.0)$ is taken for each parameter in turn. The black line is the estimate of $l(\theta)$ (or $\log(-l(\theta))$ for the first wave) obtained from additional simulator runs. The red lines show the mean and 95\% credible interval for the GP prediction of $l(\theta)$. Predictions in wave $i$ are only made in the not-implausible region $\Theta_{i-1}$. }
\label{fig:diagnostics}
\end{figure*}

The degree of conservatism of the criterion in ruling points implausible or not, is controlled by the choice of threshold $T$, and by the multiplier of $\sigma$ on the left-hand-side of the equation. For the examples considered below, the choice $T=10$ is found to provide a reasonable trade-off between accuracy and allowing sufficient space to be ruled as implausible, as $\exp(10)>10^4$, and so using the  approximation $\pi(\theta |D)=0$ if  $l(\theta) <l(\thetahat) -10$ causes only  a small error in the approximation to the posterior. 

\subsection{Sequential approach}

We aim to rule out an increasing proportion of prior input space $\Theta_0$ in a sequence of  waves. Each wave involves extending the design, determining the implausible region, running the simulator at not-implausible points, building a new GP model, and running diagnostics. We start with ensemble $\mathcal{E}_1 = \{(\theta_i, \hat{l}_M(\theta_i))_{i=1}^{N_1}\}$ where $\{\theta_i\}_{i=1}^{N_1}$ are the first $N_1$ points from a Sobol sequence dispersed to fill $\Theta_0$  as described in Section \ref{sect:design}. We denote the GP model fit  to $\mathcal{E}_1$ by $\eta_1(\cdot)$.

The design is then extended by drawing $N_2$ additional points in $\Theta_0$. For each new point in the design we  apply the implausibility criterion (\ref{eqn:implausibility}) using the mean and covariance function of $\eta_1$ to determine whether it is implausible or not, defining the not implausible region according to $\eta_1$, denoted $\Theta_1$. 
The simulator is then  run at all the new design points that were ruled to be not implausible. Collected together with the points from $\mathcal{E}_1$ that were not implausible, this gives a new ensemble $\mathcal{E}_2$. We use $\mathcal{E}_2$ to build GP model $\eta_2(\cdot)$. Note that $\eta_2$ will only give good predictions for $\theta \in {\Theta_1}$.

For the i$^{th}$ wave, we extend the design by a further $N_i$ points in $\Theta_0$. To judge whether $\theta$ is implausible, we first decide if $\theta \in \Theta_1$ using $\eta_1$, and if so, we then use $\eta_2$ to test if $\theta \in \Theta_2$, and so on. Parameter $\theta$ is only judged to be not-implausible if Equation (\ref{eqn:implausibility}) is not satisfied for all $i-1$ GP models fit in previous waves. It is necessary to use the entire sequence of GPs, as earlier GPs are only trained on the not-implausible region at that wave, and so are unable to usefully predict outside of this region, i.e., $\eta_1$ is unlikely to give poor predictions of $l(\theta)$ if $\theta \in \Theta_0  \backslash\Theta_1$.

The motivation for this sequential approach is that the size of the not implausible region $\Theta_i$ decreases with each iteration, and more importantly, the value of $l(\theta)$ is less variable in $\Theta_i$ than in $\Theta_{i-1}$. This helps the GP model to achieve superior accuracy in later waves, and in particular, the variance of the predictions decreases (as there are more design points in the region of interest). While it is possible to reduce the threshold $T$ at each wave, we keep it fixed, and instead use the decreasing uncertainty in the improved GP fits to rule out increasingly wide regions of space.

The values of $N_i$ can be chosen either in advance, or by extending the design a point at a time until the number of not-implausible design points for the next wave is sufficiently large. To determine the number of waves needed, detailed diagnostics \citep{Bastos_etal2009} can be used  to judge whether each GP model fit is satisfactory. For most problems, we have found 3 or 4 waves to be sufficient. Beyond this number, the need to iteratively use the entire sequence of GP models $\eta_1,\ldots, \eta_i$ to judge implausibility becomes increasingly burdensome. For some problems, particularly if the prior support $\Theta_0$ includes regions with very large negative $l(\theta)$ values, it may be necessary to  model $\log (-l(\theta))$ in the first wave, in order to cope with the  orders of magnitude variation seen in the log-likelihood function. In these cases, the  implausibility criterion will need to be suitably modified.


Once we have a GP model,  $\eta_I(\cdot)$ say, that accurately  predicts $l(\theta)$  within the not-implausible region $\Theta_{I-1}$, we can find the posterior distribution. We use a Metropolis-Hastings (MH) algorithm with random walk proposal. The acceptance step  iteratively uses GPs $\eta_1, \ldots, \eta_{I-1}$ to predict if the proposed parameter, $\theta'$, is implausible. If $\theta'\in \Theta_{I-1}$, then we use $\eta_I$ to predict $l(\theta')$. We use a random realization (not just the GP mean) to account for the error in the likelihood prediction, and then use the MH ratio to decide whether to accept $\theta'$ or not.
Note that the MCMC does not require any further simulator evaluations. 

\begin{figure*}
\begin{center}
\includegraphics[width=0.8\linewidth]{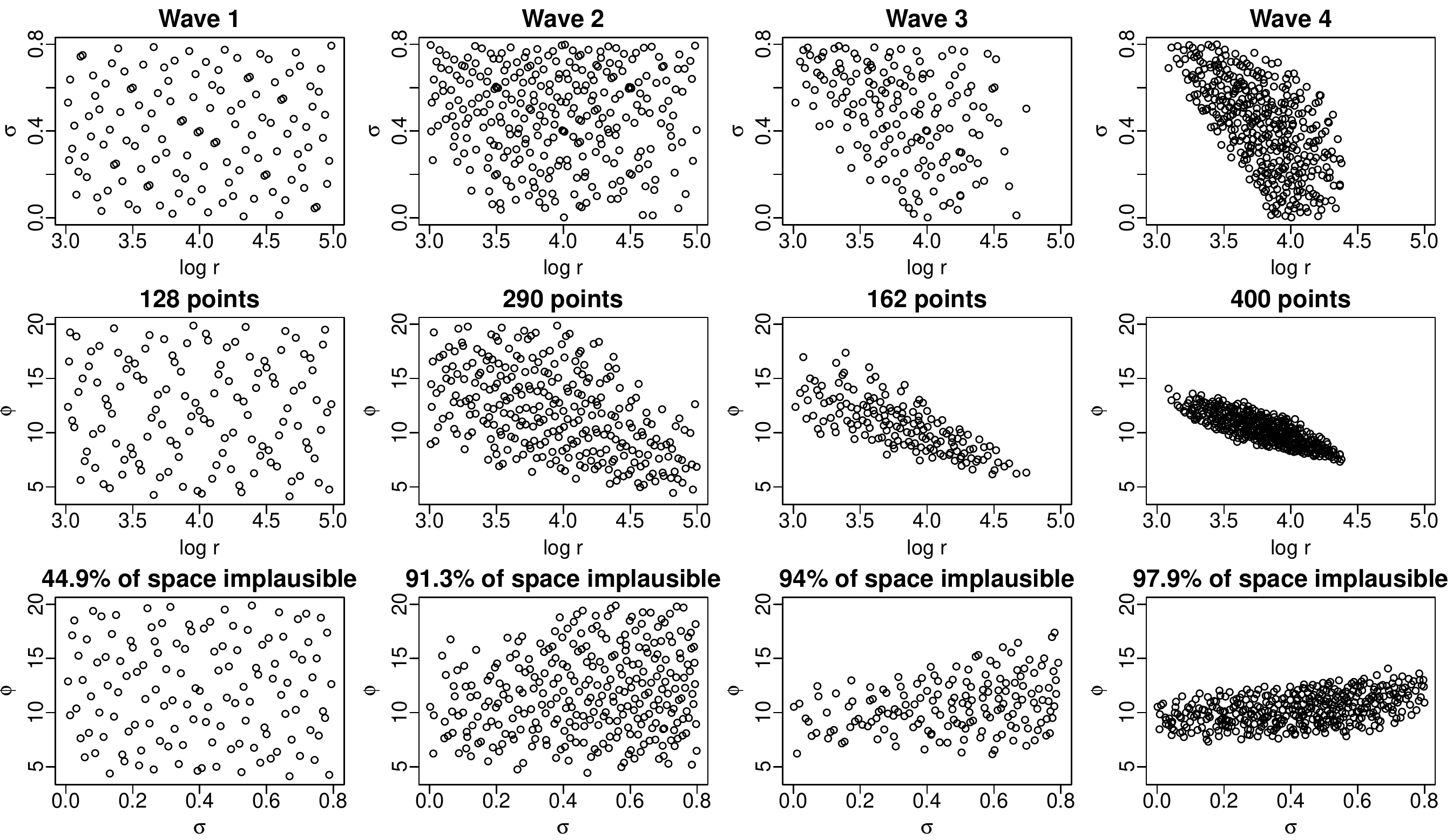} 
\end{center}
\caption{The evolution of the experimental design through four waves of history matching (one column  per wave). Each plot shows the projection of the design in the not-implausible region, $\Theta_{i-1}$,  onto two dimensions. The proportion of the prior support $\Theta_0$ ruled out increases every wave and is reported in the figure. Note that the projection of the volume onto two dimensions  acts to disguise the amount of space that has been ruled out.}
\label{fig:designs}
\end{figure*}

\section{Ricker model}\label{sect:Ricker}

The Ricker model is used in ecology to model the number of individuals in a population through time. 
Despite its mathematical simplicity, this model is often used as an exemplar of a complex model  \citep{Fearnhead_etal2012, Shestopaloff_etal2013} as it can cause the collapse of standard statistical methods due to near-chaotic dynamics \citep{Wood_2010}.
Although the model is computationally cheap, allowing the use of  expensive sampling methods such as  ABC, it is used here  to demonstrate how GP-accelerated methods can dramatically reduce the number of simulator evaluations required to find the posterior distribution.

Let $N_t$ denote the unobserved number of individuals in the population at time $t$ and $Y_t$ be the number of observed individuals. Then the Ricker model is defined by the relationships
$$N_{t+1} = rN_t {\rm e} ^{-N_t+e_t},\; Y_t \sim \operatorname{Pois}(\phi N_t), \; e_t \sim N(0, \sigma^2)$$
where the $e_t$ are independent and the $Y_t$ are conditionally independent given the $N_t$ values. 
We use prior distributions
$ \log r \sim U[3, 5], \; \sigma \sim U[0, 0.8]$, and $ \phi \sim U[4,20],$
and aim to find the posterior distribution $\pi(\theta | S(y_{1:T}))$ where $\theta= (\log r , \sigma^2, \phi)$ is the parameter vector and $y_{1:T}=(y_1, \ldots, y_T)$ is the time-series of observations.

We apply the synthetic likelihood approach used in \citet{Wood_2010}, and the GP-accelerated approach described here and compare their performance. We reduce the dimension of the data and simulator output, by using a vector of summaries $S(y_{1:T})$ which contain a collection of phase-invariant measures, such as coefficients of polynomial autoregressive models \citep[described in][]{Wood_2010}.  We use the Gaussian synthetic likelihood (Equation \ref{eqn:WoodLikelihood}) and run the simulator 500 times at each $\theta$ in the design to estimate the sample mean and covariance $ {\hat\mu}_\theta$ and ${\hat \Sigma}_\theta$. We use a simulated  dataset obtained using $\theta = (3.8,   0.3,  10.0)$. 

For the GP-accelerated inference we model $\log(-l(\theta))$ in the first wave, and $l(\theta)$ in later waves, and find that the best results are obtained using a total of four waves. We use a quadratic mean function for the GP model in the first three waves and a sixth order polynomial mean function in the final wave. After initial exploratory analysis to determine the rough shape of the log-likelihood function, we set the threshold value to be $T=3$ in the first wave (on the $\log(-l(\theta))$ scale), and $T=10$ for waves two to four, as  these thresholds were predicted to lead to negligible truncation errors.
The minimum value of the nugget variance for each GP was taken to be the variance of the estimate of $\log \piGhat(S(y_{1:T}) |\theta)$ (or $\log(-\log \piGhat(S(y_{1:T}) |\theta))$), estimated using 1000 bootstrap replicates of the sample mean and covariance matrix. Detailed diagnostic plots were used to guide these choices, a selection of which are shown in Figure \ref{fig:diagnostics}.
The accuracy of the GPs improves with each  successive wave, which is reflected in the decreasing cross-validation errors (reported in the figure).  Note that for earlier waves, it is only the ability to predict which regions are implausible that is important, not the absolute accuracy.

\begin{figure*}
\begin{center}
\includegraphics[width=0.7\linewidth]{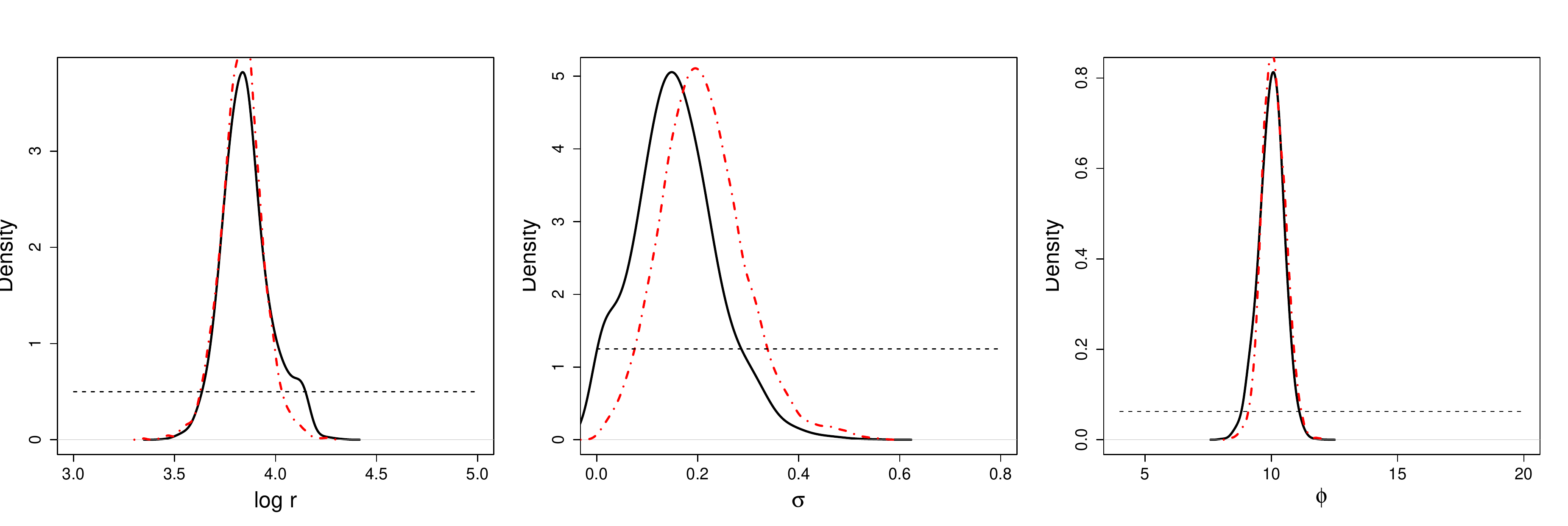} 
\end{center}
\caption{Marginal posterior distributions for the Ricker model. The solid black curve is  the posterior obtained using the synthetic likelihood approach with $10^5$ MCMC iterations. The dashed red line is the posterior obtained using the GP-accelerated approach. The constant black dotted line is the prior. The synthetic likelihood posterior required $5\times10^7$ simulator evaluations compared to only $3.5\times 10^5$ for the GP-accelerated approach.}
\label{fig:posteriors}
\end{figure*}

Through the application of the thresholds, each wave of modelling rules out an  increasing proportion of the prior support $\Theta_0$ as implausible. Figure \ref{fig:designs} shows the design used in each wave, and the proportion of space ruled out. Wave one rules out 45\% of $\Theta_0$, and by wave four, over 97\% of space has been deemed implausible.  

Figure \ref{fig:posteriors} shows the posterior distributions estimated using  the synthetic likelihood approach and the GP-accelerated approach. The GP-accelerated approach required a total of $3.5\times 10^5$ model evaluations. We ran the MCMC chain in the Wood method for $10^5$ iterations (which is probably too few), which required a total of $5\times 10^7$ simulator evaluations,  140 times more than required by the GP-accelerated approach. The  posterior distributions for $\log r$ and $\phi$ are very similar, with the exception of an additional ridge in the posterior for $\log r$ which may be genuine, or may be an artefact of not  running the synthetic likelihood MCMC for sufficiently long. The marginal posterior for $\sigma$ shows a small difference between the two methods. Estimating scale parameters is harder than estimating location parameters \citep{Cox2006}, and it is usually when estimating scale parameters that the GP-accelerated approach has been observed to have poor accuracy.

\section{Estimating species divergence times}

We now examine a model used in evolutionary biology to estimate species divergence times using the fossil record \citep{Tavare_etal2002, Wilkinson_etal09}. This model has an intractable likelihood function and has been used to demonstrate various advances in ABC methodology \citep{Marjoram_etal03, Wilkinson_2007}. The model consists of a branching process representing the unobserved phylogenetic relationships, which is randomly sampled to give a temporal pattern of fossil finds that can be compared to the known fossil record. 
We use  data on primates \citep[provided in ][]{Wilkinson_etal2011}, consisting of counts of the number of known primate species from the 14 geological epochs of the Cenozoic, denoted $\mathcal{D}=(D_1, \ldots, D_{14})$. To model these data, a non-homogeneous branching process rooted with two individuals at an assumed divergence time of $54.8 + \tau$ million years (My) ago is used (the oldest known primate fossil is 54.8My old). Informally, the parameter $\tau$ can be thought of as representing the temporal gap between the oldest primate fossil and the first primate, and is the key parameter of interest. Each species is represented as a branch in the process,
with the branching probabilities and age distribution controlled by three unknown parameters $\rho, \gamma$ and $\lambda$.
Once the branching process has been simulated, the number of species in each geological epoch are counted, giving values $\mathcal{N} = (N_1, \ldots, N_{14})$. 
The fossil data, $\mathcal{D}$,  are then assumed to be from a binomial distribution $D_i \sim \operatorname{Bin}(N_i, \alpha_i)$, with $\alpha_i = \alpha p_i$, where the $p_i$ are known sampling fractions reflecting the differing lengths of each epoch and the variation in the amount of visible rock. This gives five unknown parameters, $\theta = (\tau, \alpha, \rho, \gamma, \lambda)$, with primary interest lying in the estimation of the temporal gap $\tau$. We use uniform priors $\tau \sim U[0,100], \; \alpha \sim U[0,0.3], \; \rho \sim U[0, 0.5],\gamma \sim U[0.005, 0.015]$, and $ \lambda \sim U[0.3, 0.5]$ and try to find the posterior distribution $\pi(\theta | \CD)$.

\begin{figure*}
\begin{center}
\includegraphics[width=1\linewidth]{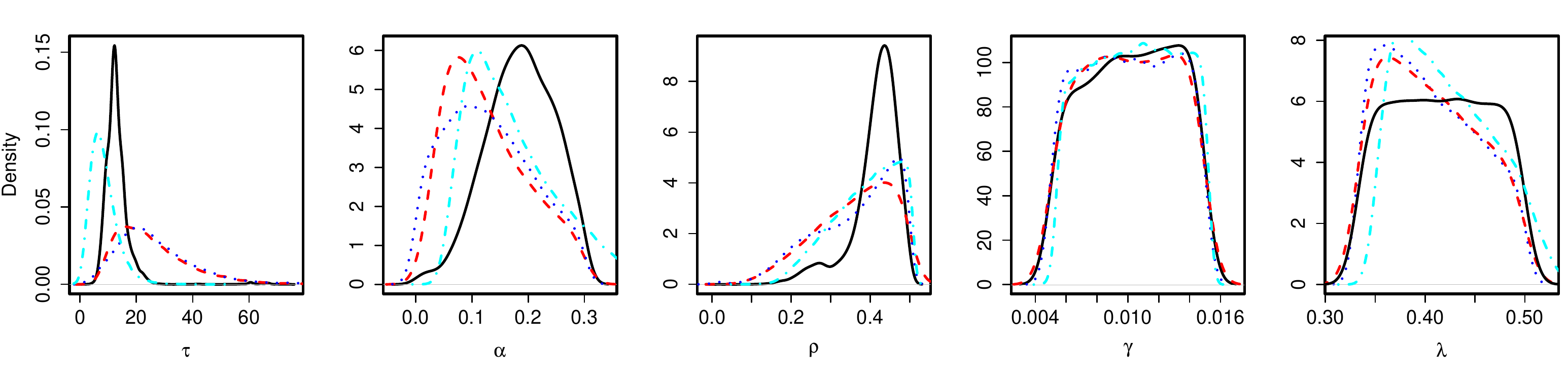} 
\end{center}
\caption{Marginal posterior distributions for the species divergence model. The solid black line shows the GP-accelerated approximation to the exact posterior. The dashed red line is from the rejection ABC approach. The dotted blue line is the GP-accelerated approximation to the ABC distribution. The dot-dash cyan line is from using a local-linear regression adjustment on the rejection ABC posterior.}
\label{fig:PrimatePosterior}
\end{figure*}


The basic rejection ABC algorithm  is simple to apply. We use the metric defined by \citet{Marjoram_etal03} with a tolerance  of $\epsilon =0.1$, and generate 2000 acceptances from the algorithm given in Section 1, which required 13.6 million simulator evaluations (results shown as dashed red lines in Figure \ref{fig:PrimatePosterior}). For the GP-accelerated approach, we estimate the likelihood for each $\theta$ in our design by 
\begin{equation}\label{eqn:ABClike}
\pi_{ABC}(\CD|\theta) \approx \frac{1}{M} \sum_{i=1}^{M}  \mathbb{I}_{\rho(D, D')\leq \epsilon} 
\end{equation}
where $\CD'$ is a simulated dataset. 
Due to the very low acceptance rate, we have to use a large value of $M$ to generate any acceptances, even when $\theta$ is near the maximum likelihood estimate. Approximately 50\% of the prior input space led to no accepted simulations after $10^4$ replicates, which we dealt with by leaving these values out of the GP fit (alternatively, we can substitute a value less than $10^{-4}$ for the ABC-likelihood). 
An additional problem, is that the estimator  of the ABC likelihood  (Equation \ref{eqn:ABClike}) has large variance,  making the training ensemble a very noisy observation of the log-likelihood surface, which necessitates the use of a large nugget term in the Gaussian process. Surprisingly,  we still find that the GP-accelerated approach is successful. Using two waves, with 128  design points in total, gives the results shown by the blue dotted lines in Figure \ref{fig:PrimatePosterior}. These results needed 1.28 million simulator evaluations, a factor of 10 fewer  compared with the rejection ABC algorithm. The accuracy of these results is good, and can be improved further by increasing the value of $M$ and by refining the GP model of the log-likelihood surface.

The use of the discontinuous acceptance kernel (the 0-1 cutoff $\mathbb{I}_{\rho(\CD, \CD') \leq \epsilon}$) in the ABC likelihood makes the log-likelihood surface difficult to model.  Using a smooth GABC acceptance kernel $\pi(\CD | \mathcal{N})$, as in Equation (\ref{eqn:piGhat}),  makes the estimate of the log-likelihood values less variable, and the surface easier to model. However, rather than present those results, we instead note that it is possible to approximate the exact posterior distribution in this case. The simulator  consists of  tractable and intractable   parts. The distribution  $\pi(\CN | \theta)$ is unknown and can only be simulated from, but the observation likelihood given the phylogeny is 
$$\pi(\CD | \CN) = \prod_{i=1}^{14}\binom{N_i}{D_i} \alpha^D_i (1-\alpha)^{N_i-D_i},$$
and can be used to obtain log-likelihood estimates. Using this expression in a likelihood based inference scheme does not work well due to the extreme variance of the values obtained when $\CN$ is drawn randomly from the model. 
Applying the GP-accelerated approach,  is successful however. We used four waves, with 32, 78, 100,  and 91 new design points in each wave, using a constant mean GP at each stage. At each design point  $10^5$ simulator replicate were used to estimate $\pi(\CD|\theta)=\BE_\CN \pi(\CD |\CN)$. The results are shown by the solid black lines in Figure \ref{fig:PrimatePosterior}. These distributions are an estimate of the  posterior  we would obtain if we could run ABC with $\epsilon =0$ (i.e., the true posterior). 
The GP-accelerated results differ from the ABC results (red line) because they approximate a different distribution, but they are consistent with our expectations. If one plots the ABC marginal posterior for $\tau$  for various values of $\epsilon$, then as $\epsilon$ decreases the posterior moves from a flat prior distribution, to an ever more peaked distribution with mass near smaller values of $\tau$. Also shown in Figure \ref{fig:PrimatePosterior} (cyan dot-dashed lines) are the posteriors obtained using the local-linear regression adjustment proposed by \citet{Beaumont_etal02}. This approach  estimates the posterior we would obtain if we could  set $\epsilon=0$ and can substantially improve the accuracy of rejection ABC. Note that the estimated posterior  is somewhere between the rejection ABC posterior and the (possibly exact) GP posterior. As this curve is an extrapolation of the trend observed as $\epsilon$ decreases, we expect the GP-ABC posterior to be more accurate.
We cannot quote any computational savings for the GP-accelerated approach, as to the best of our knowledge, it is impossible to obtain this (`exact') posterior distribution using any other approach.


\section{Conclusions}


For computationally expensive simulators, it may not be feasible to perform enough simulator evaluations to use Monte Carlo methods such as ABC at the accuracy required.  GP-accelerated methods, although adding another layer of approximation,  can provide computational savings that allow smaller tolerance values to be used in ABC algorithms, thus increasing the overall accuracy. Although the method is not universal, as it requires a degree of smoothness in the log-likelihood function, nevertheless, for a great many models this kind of approach can lead to large computational savings. 
The method requires user  supervision of the GP model building and it is important that detailed diagnostic checks are used in each wave of the GP model building. Just as poor choices of tolerance, summary and metric in ABC can lead to poor inference, similarly, poor modelling and design choices can lead to inaccuracies in the GP-ABC approach. 

Using GPs raises other computational difficulties,  as GP training has computational cost $O(N^3)$, where  $N$ is the number of training points, with complexity $O(N)$ and $O(N^2)$ for calculating the posterior mean and variance respectively. This cost means that this approach will not produce time savings if the simulator is very cheap to run.
The cost of using GPs can be reduced to $O(M^2N)$ for training (and $O(M)$ and $O(M^2)$ for prediction) by using sparse GP implementations \citep{Quinonero_etal2005}, which rely upon finding a reduced set of  $M \ll N$ carefully chosen training points  and using these to train the GP. 


Finally, the method presented here can be extended in several ways. For example, the optimal choice of the number of simulator replicates, the error induced by thresholding the likelihood,  and the location of additional design points have not been studied in detail. 
In conclusion, we have lost the guarantee of asymptotic success provided by most Monte Carlo approaches, in exchange for gaining computational tractability. Despite these drawbacks, GP-accelerated methods provide clear potential for enabling Bayesian inference in computationally expensive  simulators.


\clearpage

\bibliographystyle{abbrvnat}

{\small \bibliography{AccABC}}

\end{document}